# Unveiling the role of carbon defects in the exceptional narrowing of m-ZrO$_2$ bandgap for enhanced photoelectrochemical water splitting performance

*Ahmed H. Biby[a], Sarah A. Tolba[a] and Nageh K. Allam[a*]*

[a] Energy Materials Laboratory, School of Sciences and Engineering, The American University in Cairo, New Cairo 11835, Egypt

E-mail: Nageh.allam@aucegypt.edu

Keywords: water splitting, DFT, zirconia, defect engineering, carbon defects

The development of efficient photoelectrodes via defect engineering of wide-band gap metal oxides has been the prime focus for many years. Specifically, the effect of carbon defects in wide-band gap metal oxides on their performance in photoelectrochemical (PEC) applications raised numerous controversies and still elusive. Herein, the effect of various carbon defects in m-ZrO$_2$ was investigated using the density functional theory to probe the thermodynamic, electronic, and optical properties of the defective structures against pristine m-ZrO$_2$. The defect formation energies revealed that elevating the temperature promotes and facilitates the formation of carbon defects. Moreover, the binding energies confirmed the stability of all studied complex carbon defects. Furthermore, the band edge positions against the redox potentials of water species revealed that all the studied defective structures can serve as photoanodes for water splitting. Additionally, C$_{O3c}$ (carbon atom substituted O$_{3c}$ site) was the only defective structure that exhibited slight straddling of the redox potentials of water. Importantly, all investigated defective structures enhanced light absorption with different optical activities. Finally, C$_{O3c}$V$_{O3c}$ (carbon atom substituted O$_{3c}$ associated with O$_{3c}$ vacancy) defective m-ZrO$_2$ enjoyed low direct band gap (1.9 eV), low defect formation energy, low exciton binding energy, high mobility of charge carriers, fast charge transfer, and low recombination rate. Concurrently, its optical properties were exceptional in terms of high absorption, low reflectivity and improved static dielectric constant. Hence, the study recommends C$_{O3c}$V$_{O3c}$ defective m-ZrO$_2$ as the leading candidate to serve as a photoanode for PEC applications.

# 1. Introduction

Since the first demonstration of Solar-driven water splitting in 1972 [1], the scientific community is striving to design an efficient, durable, reliable, stable, and robust semiconductor-based photoelectrochemical (PEC) system in quest of the maximum theoretical solar-to- hydrogen (STH) efficiency of 29.7% [2]. Given that the maximum practically achieved STH efficiency is still lagging below 1.0% [2], an immense demand of persistent efforts for revolutionizing the performance of the PEC main components is needed. A typical PEC system consists of photoelectrodes, electrolyte, and separation membrane [3]. In PEC system, the adequate performance relies heavily on the photoelectrodes. Therefore, optimizing the properties of the photoelectrodes is of paramount significance towards the advancement and the commercialization of feasible cost-effective PEC systems for water splitting [2]. The key factors that dictate the quality of a photoelectrode are the energy band-gap, electron-hole recombination, stability, and the photocurrent density [2]. Metal oxides possess high photoelectrochemical stability, wide range of bandgaps and satisfactory band edge positions [4], yet they have limited light absorption, high recombination rates, poor electrical conductivity, and short hole diffusion length [2]. In this regard, defect engineering through doping is identified as an effective approach to enhance the optical and electronic properties of metal oxides. Doping improves the light absorption of the material by introducing defect states that narrow the bandgap and improves the conductivity [5]. To this end, a myriad of dopants and defects in metal oxides were explored for water splitting applications [6–15]. Among the various dopants, carbon doping in titania raised a huge controversy in literature whether it can induce visible light absorption or not. Khan et al. claimed that carbon doping in titania resulted in unprecedented efficiency of 8.35 % [16] whereas Lackner, Fujishima and Murphy criticized the study and attributed the flawed result to the erroneous measurement protocol of evading the placement of AM 1.5 filter and the underestimation of the bias voltage [17–19]. Nevertheless, numerous experimental and computational studies showed promising results of specific carbon defects in titania for water splitting applications [20–26]. Similar to titania, zirconia is a wide-gap metal oxide with distinctive properties such as high melting-point, chemical inertness and high corrosion tolerance [27] that render it advantageous for a plethora of applications [28-49]. However, the bandgap of zirconia needs to be optimized to enhance its photocatalytic activity. Unlike titania, the use of zirconia in water splitting systems has been poorly addressed in literature. A few studies shed the light on the effect of various dopants on the performance of zirconia in solar-driven hydrogen production systems[50–52]. Although carbon-doped zirconia was briefly investigated in the literature [53–55] the debatable effect of carbon doping in zirconia was neither treated nor unriddled systematically and comprehensively for water splitting applications. The first-principles calculations proved to be credible and efficient in diverse applications [56–64]. Therefore, Herein, an in-depth first-principles study was conducted to encompass the viable scenarios for doping monoclinic zirconia with carbon and to scrutinize the effect of carbon doping on its performance metrics as a photoelectrode for water splitting. Particularly, our study investigated the thermodynamic, electronic, and optical properties of the studied defects to report the best candidate defective structure for the photoelectrodes of PEC systems.

## 2. Methods

In our study, CASTEP code [65] was utilized for the spin-polarized density functional theory (DFT) calculations. For modelling the electron-electron interaction, the PBE-GGA (Perdew-Burke-Ernzerhof of Generalized Gradient Approximation) [66] was chosen. The Ultrasoft-pseudopotential [67] was adopted to account for the electron-ion interaction. In the framework of plane-wave basis set, the Kohn-Sham wave functions of electrons were expanded up to 380 eV. In consideration of the irreducible Brillouin zone, the Monkhorst-Pack scheme [68] was implemented to sample the k-points where a mesh of (5x5x5) and (2x5x5) (0.04 spacing of points) k-points were constructed for the single unit cell and the supercell, respectively. Both were used for geometry optimization calculations and generation of density of states (DOS) profile. In order to remedy the self-interaction error in DFT [69], the Hubbard correction approach [70] was employed to describe the energy band gap adequately. Furthermore, Hubbard correction was exploited to attain proper positions of the defect states within the band gap and to improve the accuracy of the defect formation energy calculations [71]. In this regard, the applied Hubbard U parameters, were 4 eV to the 4d orbitals of Zr and 4 eV on the 2p orbitals of O to maintain the nature of Zr-O covalent bond as explained in ref. 72. Concerning the geometry optimization convergence criteria, the mean Hellmann-Feynman force was set to 0.01 eV/A while the maximum displacement tolerances, maximum stress, and energy change were adjusted to $5.0 \times 10^{-4}$, 0.02 GPa, and $5.0 \times 10^{-6}$ eV/atom, respectively. In pursuance of benchmarking of the structures and computational setup, the coordinates of the atoms and the lattice parameters of m-$ZrO_2$ were acquired from Purohit et al. work [73] and the calculated parameters were compared to their counterparts in literature. After structural relaxations, the obtained energy band gap and lattice parameters matched the previously reported experimental and theoretical values [74]. The values were $E_g$=5.14 eV, a =5.24 Å, b=5.20 Å, and c=5.40 Å, respectively. Hence, the computational setup was accredited to be inherited in the optimization of the defected structures. For the sake of studying all the possible carbon defected structures, a (2x1x1) supercell including 24 atoms was established to embrace the different defect concentrations and to minimize the electrostatic interaction with the periodic images. Additionally, two defect concentrations were taken into consideration to represent the low and high concentrations specifically; 0.125 and/or 0.25 nD/nZr (D=defect) where one and/or two defects were introduced to the supercell. It is noteworthy that m-$ZrO_2$ has two inequivalent oxygen sites of different binding energies namely; three-coordinated ($O_{3c}$) and four-coordinated ($O_{4c}$). Initially, the oxygen-deficient (reduced) $ZrO_2$ was simulated in two scenarios with 0.125 and 0.25 $V_O$ defects following the approach of Sinhamahapatra et al. [75]. Afterwards, interstitial, substitutional, and complex carbon defects were incorporated into the pristine and the reduced $ZrO_2$ as described by the defects equations in **Table 1**. Starting from the optimized pristine structure, the following structures were modelled: (i) $C_{O3c}$ (carbon atom substituted $O_{3c}$), (ii) $C_{O4c}$ (carbon atom substituted $O_{4c}$), (iii) $C_{O3c}$ - $C_{O3c}$ (two adjacent bonded $C_{O3c}$), (iv) $C_{O3c}=O_{3c}$ ($C_{O3c}$ bonded with adjacent $O_{3c}$) and (v) $C_i$ (carbon atom placed between two adjacent $O_{3c}$ and $O_{4c}$). With respect to the structures which started from the optimized reduced $ZrO_2$, the subsequent configurations were studied: (vi) $C_{Vo3c}$ (carbon atom occupied $O_{3c}$ vacancy), (vii) $C_{Vo4c}$ (carbon atom occupied $O_{4c}$ vacancy), (viii) $V_{O4}C_{O3c}$ ($V_{O4c}$ associated with $C_{O3c}$), and (ix) $V_{O4c}C_i$ ($V_{O4c}$ associated with $C_i$). Other calculations started from m-$ZrO_2$ defected with $C_{O3c}$ such as: (x) $C_{O3c}V_{O3c}$ ($C_{O3c}$ associated with $V_{O3c}$) (xi) $C_{O3c}V_{O4c}$ ($C_{O3c}$ associated with $V_{O4c}$) while the last structure started from m-$ZrO_2$ with interstitial carbon defect; (xii) $C_iV_{O4c}$ ($C_i$ associated with $V_{O4c}$). It is obvious that (i) is identical to (vi), (ix) is identical to (xii) and (viii) is identical to (xi), albeit the starting structures are different. The defect formation energy was calculated to assess the relative difficulty of introducing the defects into the structure. Apart from this, binding energy was also computed for complex (associated) defects as an essential thermodynamic stability indicator against their decomposition into their rudimentary defects [76]. The formation energy of a given neutral defect (D) is denoted by $E_D^f$ and can be defined as:

$$E_D^f = E_{defected} - E_{pristine} + \sum_i \Delta n_i \mu_i \quad (1)$$

where $E_{defected}$ and $E_{pristine}$ represent the calculated total energies of the defected supercell and the pristine supercell, respectively, $\Delta n_i$ is the difference between the number of a given species i in the pristine supercell and the number of the same species in the defected supercell. Eventually, $\mu_i$ symbolizes the chemical potential of a given species i. The chemical potential calculations depend on the thermodynamic reservoir of the added/removed species and the experimental growth conditions that may be oxygen-rich or zirconium-rich as extreme conditions. For oxygen-rich (zirconium-poor) condition, $\mu_O$ can be obtained from ground state energy of oxygen molecule whereas $\mu_{Zr}$ and $\mu_C$ can be calculated from $\mu_{Zr}= \mu_{ZrO2} -2\mu_O$ and $\mu_C= \mu_{CO2}-2\mu_O$, respectively. Under the zirconium-rich (oxygen-poor) condition, $\mu_{Zr}$ can be acquired from ground state energy of single zirconium atom in bulk zirconium while $\mu_O$ can be provided by $\mu_O = (\mu_{ZrO2} - \mu_{Zr})/2$ then $\mu_C$ can be estimated via $\mu_C = \mu_{CO2} - 2\mu_O$. On the other hand, the binding energy of a given complex C is expressed by $E_b$ and its formula is:

$$E_b = E_C^f - \sum_l E_D^f \qquad (2)$$

where $E_C^f$ and $E_D^f$ are the formation energies of a given complex and the formation energy of a defect that constitutes the complex C, respectively. The number of l species equals the number of defects composing the given complex.

## 3. Results and Discussion

The impact of the various carbon defects in pristine and reduced m-$ZrO_2$ was investigated in light of the modification in electronic and optical properties.

### 3.1 Defect Energetics

Before delving into the electronic and optical properties, analysing the relative thermodynamic stability of the defects in the doped structures is crucial through the comparison of the formation energies of the defects and the binding energies of the complex defects. The formation energies of the studied structures at 100 and 1000 K are listed in **Table 1**. Generally, from a thermodynamic point of view, formation energies of carbon defects are more favourable under oxygen poor conditions at both high and low temperatures. Thus, the discussion would focus only on poor oxygen conditions. For carbon-substituting-oxygen structures, it was found that carbon substituting $O_{3c}$ atom is energetically more stable than substituting $O_{4c}$ atom. Concerning carbon interstitial defect, it was observed to have the highest formation energy. Regarding complex defects formed in pristine m-$ZrO_2$, it was noticed that forming a complex of two adjacent $C_{O3c}$ defects with a single bond between the two carbon atoms is more favourable to exist rather than $C_{O3c}$ defect alone, where $C_{O3c}$ - $C_{O3c}$ (case iii) has the lowest formation energy at 1000 K. In addition, the formation of double bond between $C_{O3c}$ defect and its adjacent $O_{3c}$ atom was identified as more stable complex than the $C_{O3c}$ defect alone with a difference in their formation energies reaching -0.12 eV.

Upon addressing the effects of carbon defects on inducing oxygen vacancies, it was reported in our previous study that formation of $V_{O4c}$ is prevalent over formation of $V_{O3c}$ [51]. However, in this study, it was found that the presence of $C_{O3c}$ atom initiates the formation of adjacent $V_{O3c}$ that is more likely to occur than the formation of $V_{O4c}$ with a difference in their formation energies of -1.75 eV. This may be attributed to the fact that $V_{O3c}$ bonding environment is more preferred for carbon since carbon can form either a double bond with $O_{3c}$ (iv case) or a complex with adjacent $C_{O3c}$ (case iii). Also, the formation energy of $V_{O4c}$ with a prior existence of $C_i$ is -2.5 eV, which renders interstitial carbon defects effectively induce the formation of oxygen vacancies within m-$ZrO_2$, exothermally, in contrast to the formation of oxygen vacancies in m-$ZrO_2$ that have positive formation energies. We also report that the formation of $C_i$ is easier in case of oxygen-deficient m-$ZrO_2$ than in pristine m-$ZrO_2$ crystal. Moreover, calculations indicated that $C_{VO3c}$ is more thermodynamically possible than $C_{O3c}$. Finally, the formation of $C_{O3c}$ adjacent to $V_{O4c}$ (case viii) was found to be more likely to occur than the formation of $C_{O3c}$ in pristine m-$ZrO_2$. Generally, the considered fabrication conditions (oxygen-rich or -poor) and the temperature have a significant impact on the formation energy of the defected structures. In summary, we can conclude from **Table 1** that elevating the temperature promotes and facilitates the formation of carbon defects and complex defects, especially in O-poor condition where the percentage of decrease in formation energies from 100 to 1000 K is significant.

**Table 1** Formation energy ($E_D^f$) of all defected m-ZrO$_2$ structures.

| Initial structure | Defect | Formation Energy ($E_D^f$) (eV) | | | | Defect Equations |
|---|---|---|---|---|---|---|
| | | O-rich | | O-poor | | |
| | | 100 K | 1000 K | 100 K | 1000 K | |
| m-ZrO$_2$ | (i) C$_{O3c}$ | 14.46 | 12.95 | 2.45 | 0.94 | $CO_2(g) + O_{O3c}^x \rightarrow C_{O3c}'''' + \frac{3}{2}O_2(g) + 2h^\bullet$ |
| | (ii) C$_{O4c}$ | 14.67 | 13.16 | 2.66 | 1.15 | $CO_2(g) + O_{O4c}^x \rightarrow C_{O4c}'''' + \frac{3}{2}O_2(g) + 2h^\bullet$ |
| | (iii) C$_{O3c}$ - C$_{O3c}$ | 23.82 | 20.80 | -0.20 | -3.22 | $2CO_2(g) + 2O_{O3c}^x \rightarrow 2C_{O3c}'''' + 3O_2(g) + 4h^\bullet$ |
| | (iv) C$_{O3c}$ =O$_{3c}$ | 14.34 | 12.83 | 2.33 | 0.82 | $CO_2(g) + O_{O3c}^x \rightarrow C_{O3c}'''' + \frac{3}{2}O_2(g) + 2h^\bullet$ |
| | (v) C$_i$ | 12.00 | 11.50 | 3.99 | 3.50 | $CO_2(g) + V_i^x \rightarrow C_i'''' + O_2(g) + 4h^\bullet$ |
| Oxygen-deficient m-ZrO$_2$ | (vi) C$_{VO3c}$ | 8.55 | 8.07 | 0.55 | 0.07 | $CO_2(g) + V_{O3c}'' \rightarrow C_{O3c}'''' + \frac{1}{2}O_2(g) + 2h^\bullet$ |
| | (vii) C$_{VO4c}$ | 8.66 | 8.17 | 0.65 | 0.17 | $CO_2(g) + V_{O4c}'' \rightarrow C_{O4c}'''' + \frac{1}{2}O_2(g) + 2h^\bullet$ |
| | (viii) V$_{O4c}$ C$_{O3c}$ | 13.16 | 11.65 | 1.16 | -0.35 | $V_{O4c}'' + CO_2(g) + O_{O3c}^x \rightarrow V_{O4c}C_{O4c}'''' + \frac{3}{2}O_2(g) + 2h^\bullet$ |
| | (ix) V$_{O4c}$ C$_i$ | 8.50 | 8.03 | 0.51 | 0.03 | $V_{O4c}'' + CO_2(g) + V_i^x \rightarrow V_{O4c}C_i'''' + O_2(g) + 2h^\bullet$ |
| C$_{O3c}$ defected m-ZrO$_2$ | (x) C$_{O3c}$ V$_{O3c}$ | 2.97 | 1.94 | -1.04 | -2.06 | $CO_2(g) + O_{O3c}^x + V_{O3c}'' \rightarrow C_{O3c}V_{O3c}'''' + \frac{3}{2}O_2(g) + 2h^\bullet$ |
| | (xi) C$_{O3c}$ V$_{O4c}$ | 4.70 | 3.69 | 0.71 | -0.31 | $CO_2(g) + O_{O3c}^x + V_{O4c}'' \rightarrow C_{O3c}V_{O4c}'''' + \frac{3}{2}O_2(g) + 2h^\bullet$ |
| C$_i$ defected m-ZrO$_2$ | (xii) C$_i$ V$_{O4c}$ | 2.53 | 1.51 | -1.47 | -2.50 | $CO_2(g) + V_i^x + V_{O4c}'' \rightarrow C_iV_{O4c}'''' + \frac{3}{2}O_2(g) + 2h^\bullet$ |

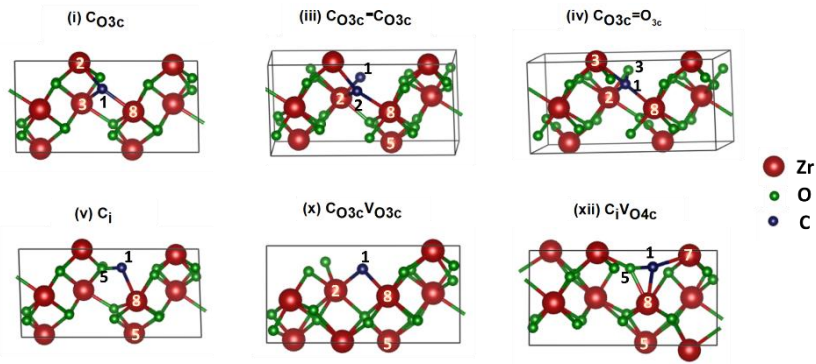

**Figure 1** Crystals structures of the short-listed defected structures.

The binding energies presented in **Table 2** reflect the thermodynamic stability of the studied complexes against dissociation over the wide range of temperature variation and regardless of the considered experimental growth conditions (O-rich/poor). Additionally, binding energies of complex defects depend on the pathway of their formation. For example, $V_{O4c}C_i$ and $V_{O4c}C_{O3c}$ have more positive binding energies than $C_i V_{O4c}$ and $V_{O4c} C_{3c}$, respectively, although the final defected structures are the same.

**Table 2** Binding energies ($E_b$) of the complex defects.

| Initial structure | Defect | Binding Energy ($E_b$) (eV) | | | |
|---|---|---|---|---|---|
| | | O-rich | | O-poor | |
| | | 100 K | 1000 K | 100 K | 1000 K |
| m-ZrO$_2$ | (iii) $C_{O3c}$ - $C_{O3c}$ | -5.10 | -5.10 | -5.10 | -5.08 |
| Oxygen-deficient m-ZrO$_2$ | (ix) $V_{O4c} C_i$ | -9.40 | -8.35 | -6.45 | -5.42 |
| | (viii) $V_{O4c} C_{O3c}$ | -7.20 | -6.18 | -4.26 | -3.24 |
| $C_{O3c}$ defected m-ZrO$_2$ | (x) $C_{O3c} V_{O3c}$ | -17.29 | -15.79 | -5.29 | -3.77 |
| $C_i$ defected m-ZrO$_2$ | (xi) $C_{O3c} V_{O4c}$ | -15.66 | -14.14 | -4.71 | -3.20 |
| | (xii) $C_i V_{O4c}$ | -15.37 | -14.87 | -8.43 | -7.95 |

In this study, the discussion will be focused on the subsequent defected structures: (i), (iii), (iv), (x), (xii), and (v), for the following reasons. First, the defected structures (i), (ix), and (viii) are identical to (vi), (xii), and (xi), respectively. Second, the $E_D^f$ of (i) $C_{O3c}$ < (ii) $C_{O4c}$, and the $E_D^f$ of (x) $C_{O3c} V_{O3c}$ < (xi) $C_{O3c} V_{O4c}$. Third, (iii) and (iv) were the most thermodynamically favourable defects in pristine m-ZrO$_2$. Lastly, (v) would be also examined for the sake of comprehensive scanning of carbon defects in m-ZrO$_2$. The structural parameters of the selected defected structures are listed in **Table 3** and depicted in **Figure 1**. It was observed that the defected structures introduced distortions to the m-ZrO$_2$ crystal by increasing the lattice volume and modifying the angles. It is also worth mentioning that the lowest increment in the lattice volume was observed in the case of $C_{O3c}V_{O3c}$ defected structure while $C_i$ defected structure showed the highest volume increase.

Table 3 Lattice parameters for the short-listed defected structures.

| Structure | Lattice Parameters | | | | | | Volume (Å)³ |
|---|---|---|---|---|---|---|---|
| | a (Å) | b (Å) | c (Å) | α (°) | β (°) | γ (°) | |
| m-ZrO$_2$ | 10.49 | 5.22 | 5.41 | 90.00 | 99.27 | 90.00 | 292.00 |
| (i) C$_{O3c}$ | 10.61 | 5.25 | 5.42 | 89.35 | 100.01 | 90.98 | 297.03 |
| (iii) C$_{O3c}$-C$_{O3c}$ | 10.66 | 5.28 | 5.35 | 91.66 | 98.68 | 92.30 | 297.17 |
| (iv) C$_{O3c}$=O$_{3c}$ | 10.57 | 5.28 | 5.39 | 89.19 | 98.12 | 89.92 | 297.65 |
| (v) C$_i$ | 10.89 | 5.26 | 5.50 | 87.64 | 101.48 | 92.29 | 308.39 |
| (x) C$_{O3c}$V$_{O3c}$ | 10.50 | 5.29 | 5.31 | 87.72 | 94.71 | 90.27 | 293.68 |
| (xii) C$_i$V$_{O4c}$ | 10.38 | 5.27 | 5.50 | 88.72 | 98.11 | 90.65 | 297.26 |

**3.2 Electronic Properties**

Analysing the atomic orbitals quantitatively and qualitatively is instrumental for deep understanding of the electronic structure that fully dictates electronic properties of pristine and defected
m-ZrO$_2$. In this endeavour, the partial density of states (PDOS) shown in **Figure 2** as well as the charge populations and population ionicity index (P$_i$) expressed in **eq. 3** [77] have been probed to explore the entity of the defect states and the nature of the bonding environment of the studied defective structures as illustrated in **Table 4**. Additionally, light would be shed on the role and the effect of defect states in modifying bandgap and the effective masses of the charge carriers for the short-listed defective structures as listed in **Table 5**. Also, to pursue the intricate criteria of photoelectrodes, the band edges alignment has been examined against the NHE potential as depicted in **Figure 3**.

$$P_i = 1 - e^{-\left|\frac{P_c - P}{P}\right|} \quad (3)$$

where P, and P$_C$ are the bond population according to Mulliken charge population and the bond population of pure covalent bond (P$_c$ =1), respectively. P$_i$ ranges between 0 and 1 representing pure covalent and pure ionic bonds, respectively.

Upon analysing the PDOS and electronic band structure, it was found that the incorporation of carbon atoms within the m-$ZrO_2$ crystal ($E_g$ = 5.14 eV) introduced intermediate band(s) that resulted in emergence of new bandgap(s) along with the original bandgap (VBM to CBM). Importantly, with the aid of Tauc analysis of optical absorption coefficients of the studied structures, the dominant optical transition was identified to define the dominant new bandgap. Herein, the dominant new bandgap would be endorsed as the bandgap for the upcoming analysis and discussion; since the emerging new bandgap is more favourable for solar energy absorption than original bandgap of pristine m-$ZrO_2$. In case of (i) and (iii), the carbon defects initiated intermediate sub-conduction band while the rest cases, the carbon defects initiated intermediate sub-valence band. Moreover, in case of (i) and (V), intermediate defect bands laid amid of the new bandgap, which act as deep localized defect states (trap states) that cause an increase of the recombination rate of charge carriers. By inspection of **Figure 2**, (iv), (v), (x), and (xii) defective structures showed excellent bandgap narrowing down to nearly 2 eV. On the other hand, (i) and (iii) offered mild bandgap narrowing to reach 2.95 and 2.35 eV, respectively. From the analysis of PDOS in **Figure 2** and the data in **Table 4**, it was noticed that in (i), (v), and (xii), the defect states are splitted, which may be attributed to the variation of the bonding environment around $p_x$, $p_y$, and $p_z$ orbitals of the defect species that led to degeneracy breakage. In contrast to the latter case, the defect states in (iii) and (x) showed broadened peaks that can be explained in terms of the covalent bonds formed between the two carbon atoms in (iii) and the increased covalency nature in (x) between carbon and the neighbouring zirconium atoms. The left shift of the defect states in (x) and the right shift in (iii) can be credited to the availability of electrons. Thus, in (x) carbon atom is associated with a vacancy that has dangling bonds with extra electrons that can be used to form new bonds with the surrounding atoms, which is reflected in the bond population as shown in **Table 4**. Consequently, this causes the stabilization of carbon orbitals accompanied by the decrease in their energies (left shift), on the contrary of (iii) where carbon atom is bonded with another carbon atom that has the same electronegativity (unlike (x)); depriving each other from withdrawing enough electrons - for bonding - from the host crystal since the available electrons are distributed among them in contrast to (x) as illustrated in their atomic charges in **Table 4**. This rendered the carbon orbitals in (iii) less stable causing the right shift in the PDOS compared to (x). Concerning (iv), the established partial-ionic bond between carbon and oxygen atoms induced polarization that slightly incremented the density of the first peak and merged the second defect peak in (i) into the conduction band causing the narrowing of the bandgap to 2.05 eV. Regarding (v), the interstitial carbon experienced high repulsive forces due to the congested atomic environment that can be reflected in the fact that it had the highest formation energy in O-poor conditions as presented in **Table 1**. Therefore, in (xii) the presence of a vacancy nearby the interstitial carbon relieved the repulsive forces experienced by the carbon orbitals; causing the minimization of the separation between the two defect states relative to (v). Similar to (x), the dangling bonds in $V_{O4c}$ offered electrons for the carbon atom leading to the left shift of the defect state peaks in comparison with (v) and also increased the electronic occupation of carbon (decreased the atomic charge) from -0.1 to -0.53 a.u.

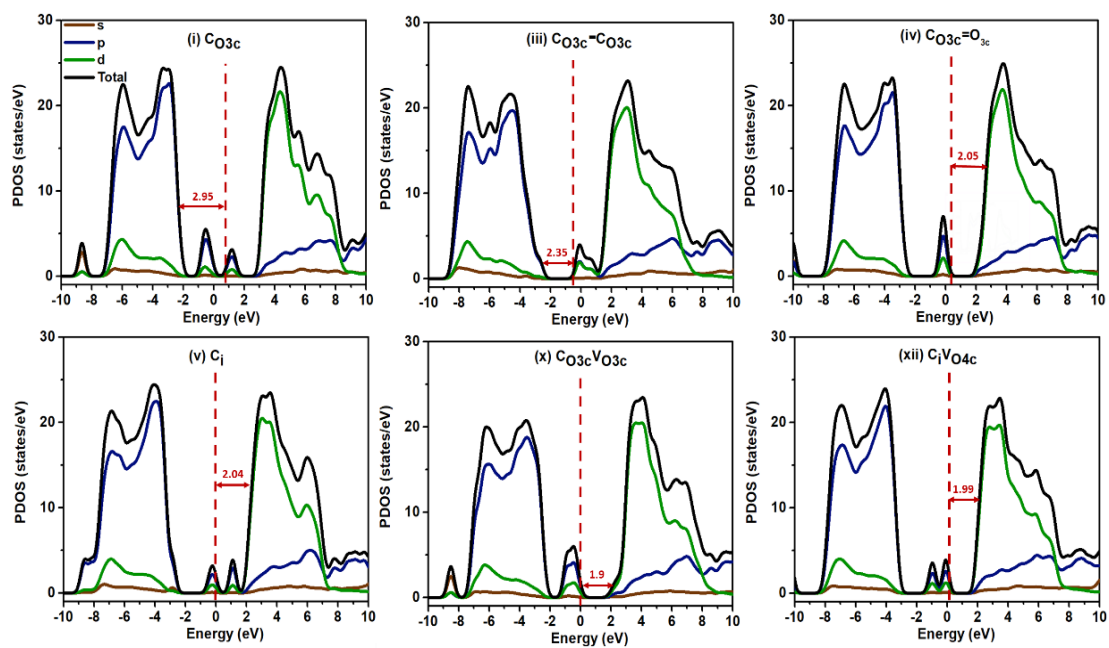

**Figure 2** Partial density of states for the short-listed m-$ZrO_2$ defective structures.

**Table 4** Atomic charge, bond length, Mulliken population, ionicity index, and bandgap for the short-listed defective structures.

| Structure | Atomic charge (a.u) | Bond | Bond length (Å) | Bond population (P) | Ionicity index ($P_i$) | Bandgap (eV) |
|---|---|---|---|---|---|---|
| **(i) $C_{O3c}$** | -0.54 | $C_1$ -- $Zr_2$ | 2.30 | 0.29 | 0.91 | 2.95 (indirect) |
| | | $C_1$ -- $Zr_3$ | 2.22 | 0.46 | 0.69 | |
| | | $C_1$ -- $Zr_8$ | 2.22 | 0.40 | 0.77 | |
| **(iii) $C_{O3c}$-$C_{O3c}$** | $C_1$ = -0.68 | $C_1$ -- $C_2$ | 1.35 | 1.49 | 0.28 | 2.35 (indirect) |
| | $C_2$ = -0.58 | $C_1$ -- $Zr_2$ | 2.27 | 0.52 | 0.60 | |
| | | $C_2$ -- $Zr_2$ | 2.54 | 0.25 | 0.95 | |
| | | $C_1$ -- $Zr_5$ | 2.42 | 0.25 | 0.95 | |
| | | $C_2$ -- $Zr_8$ | 2.30 | 0.19 | 0.98 | |
| **(iv) $C_{O3c}$=$O_{3c}$** | O = -0.69 | $C_1$ -- $O_3$ | 1.49 | 0.46 | 0.69 | 2.05 (indirect) |
| | C = -0.56 | $C_1$ -- $Zr_2$ | 2.24 | 0.05 | 1.00 | |
| | | $C_1$ -- $Zr_3$ | 2.16 | 0.62 | 0.45 | |
| | | $C_1$ -- $Zr_8$ | 2.21 | 0.47 | 0.68 | |
| **(v) $C_i$** | -0.1 | $C_1$ -- $O_5$ | 1.29 | 0.74 | 0.30 | 2.04 (indirect) |
| | | $C_1$ -- $Zr_5$ | 2.41 | 0.35 | 0.84 | |
| | | $C_1$ -- $Zr_8$ | 2.25 | 0.37 | 0.82 | |
| **(x) $C_{O3c}V_{O3c}$** | -0.92 | $C_1$ -- $Zr_2$ | 2.21 | 0.48 | 0.66 | 1.90 (direct) |
| | | $C_1$ -- $Zr_3$ | 2.09 | 0.61 | 0.47 | |
| | | $C_1$ -- $Zr_5$ | 2.36 | 0.41 | 0.76 | |
| | | $C_1$ -- $Zr_8$ | 2.22 | 0.47 | 0.68 | |
| **(xii) $C_iV_{O4c}$** | -0.53 | $C_1$ -- $O_5$ | 1.48 | 0.49 | 0.65 | 1.99 (direct) |
| | | $C_1$ -- $Zr_5$ | 2.27 | 0.44 | 0.72 | |
| | | $C_1$ -- $Zr_7$ | 2.29 | 0.59 | 0.50 | |
| | | $C_1$ -- $Zr_8$ | 2.19 | 0.19 | 0.99 | |

### 3.2.1 Band edge position

For photoelectrochemical water splitting applications, it is essential to account for the relative positions of the bands of the semiconductor to the redox potentials of water species (oxygen and hydrogen). Hence, for overall water splitting with no bias, the conduction band minimum (CBM) and the valence band maximum (VBM) of the semiconductor are required to straddle the reduction potential of water and the oxidation potential of water, respectively. However, satisfying a part of the aforementioned condition is sufficient to qualify the semiconductor to work as a photoanode if the VBM condition is fulfilled while it can serve as a photocathode if the CBM condition is fulfilled. With respect to the normal hydrogen electrode (NHE), the relative displacements between the CBM and the VBM of the short-listed structures were computed given their absolute electronegativities as shown in the following equations [51]:

$$E_{VBM} = \chi - E_e + \frac{1}{2} E_g \quad (4)$$

$$E_{CBM} = E_{VBM} - E_g \quad (5)$$

where $E_{VBM}$ and $E_{CBM}$ represent the VBM and CBM potentials, respectively, $\chi$ represents the absolute electronegativity of a given structure and it is evaluated through the calculation of the geometric mean of the electronegativities of the isolated constituting atoms of a given structure, and $E_e$ (4.5 eV) represents the energy of free electrons according to the hydrogen scale. The band edge dispositions for the short-listed m-$ZrO_2$ defective structures is depicted in **Figure 3** against the redox potentials for water species. From the figure, it is clear that all the defective structures can work as photoanodes. However, in terms of optical activity, (x) $C_{O3c}V_{O3c}$ is the best candidate as it has the lowest bandgap (1.9 eV). It is also important to clarify that (i) $C_{O3c}$ is the only defective structure that straddles the redox potentials of both water species; thus, it can theoretically serve as both photoanode and photocathode undergoing overall water splitting.

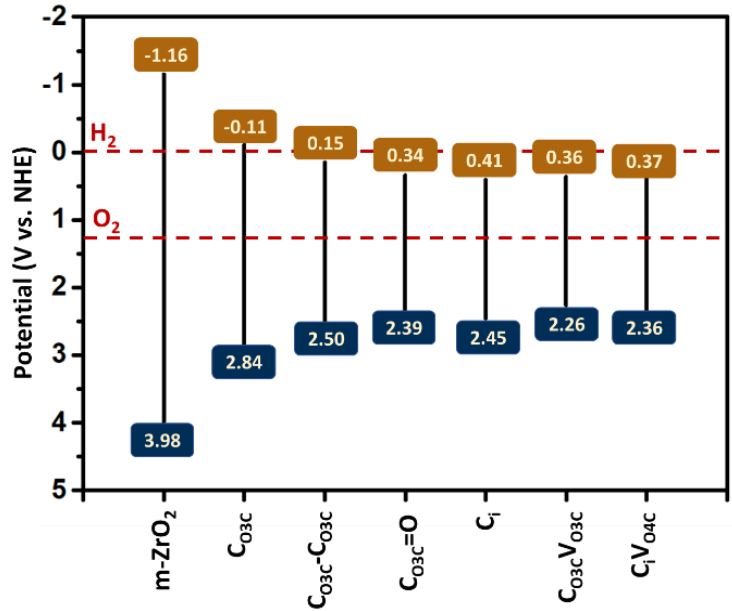

**Figure 3** Band edge dispositions for the short-listed m-ZrO$_2$ defective structures relative to the redox potentials for water species.

### 3.2.2 Effective Mass and Exciton Binding Energy

The photocatalytic activity relies upon the readiness of the electron-hole pair (exciton) generation. In this regard, the effective mass ($m^*$) of the photogenerated charge carriers is substantial; as it describes the easiness of the motion of the photogenerated carriers within the semiconductor crystal relative to the mass of free electron as specified in **eq. (6)** [51]. Also, the exciton binding energy ($E_{ex}$ in eV) is significant. Thus, it quantifies the energy needed to dissociate an electron-hole pair at its lowest energy state as formulated in **eq. (7)** [78].

$$m^* = \pm \hbar^2 \left(\frac{d^2 E_k}{dk^2}\right)^{-1} \quad (6)$$

$$E_{ex} \approx 13.56 \frac{m^\dagger}{m_e \varepsilon} \quad (7)$$

where $m_h^*$ and $m_e^*$ (kg) represent the effective masses of holes and electrons; respectively, $\hbar$ represents reduced Planck's constant, $E_k$ represents the energy corresponding to the wave vector $k$, $m^\dagger$ represents the reduced effective mass of the exciton ($1/m^\dagger = 1/m_h^* + 1/m_e^*$), $m_e$ is the rest mass of an electron, and ε is the dielectric constant. Note that D in **Table 5** represents the ratio between $m_e^*$ and $m_h^*$ and it reflects the charge recombination rate. To this end, a large difference is required between the effective masses of the photogenerated electrons and holes to secure the spatial separation between them ensuring low charge carrier recombination rates [79], thus enhancing the photocatalytic efficiency. For the short-listed structures, (iii), (x), and (xii) gave the best D values indicating low recombination rates. However, (x) $C_{O3c}V_{O3c}$ demonstrated the lowest $m_e^*$; resulting in high mobility and fast charge transfer in comparison to all studied structures including m-ZrO$_2$. For efficient dissociation of excitons at room temperature, $E_{ex}$ is necessary to be lower than k$_B$T ~ 25 meV [80]. (x) $C_{O3c}V_{O3c}$ provided the highest reduction in $E_{ex}$ (91 meV) relative to m-ZrO$_2$ (169 meV). Even though 91 meV is greater than the required $E_{ex}$ (25 meV), the DFT calculations were employed at 0 K thus, the required $E_{ex}$ is predicted to be lower than 91 meV and probably lower than 25 meV at room temperature.

### 3.3 Optical Properties

Investigation of the optical properties for candidate materials of electrodes for photoelectrochemical water splitting is a paramount decisive step for such an intricate optoelectronic application.

**Table 5** Effective masses and exciton binding energy for the photogenerated charge carriers of the short-listed defective structures.

| Structure | | $m_h^*/m_e$ | | $m_e^*/m_e$ | | D | E$_{ex}$ (meV) |
|---|---|---|---|---|---|---|---|
| m-ZrO$_2$ | Direction | G→F | G→Z | G→F | G→Z | 2.72 | 169 |
| | Calculation | 0.33 | 0.25 | 0.38 | 1.20 | | |
| | Average | 0.29 | | 0.79 | | | |
| (i) C$_{O3c}$ | Direction | G→F | G→Z | Z→Q | Z→G | 2.69 | 179 |
| | Calculation | 0.42 | 0.32 | 1.77 | 0.22 | | |
| | Average | 0.37 | | 1.00 | | | |
| (iii) C$_{O3c}$-C$_{O3c}$ | Direction | G→F | G→Z | Z→Q | Z→G | 9.56 | 655 |
| | Calculation | 0.2 | 2.97 | 29.72 | 0.59 | | |
| | Average | 1.59 | | 15.16 | | | |
| (iv) C$_{O3c}$=O$_{3c}$ | Direction | G→F | G→Z | Z→Q | Z→G | 0.32 | 176 |
| | Calculation | 1.22 | 1.25 | 0.50 | 0.28 | | |
| | Average | 1.24 | | 0.39 | | | |
| **(v) $C_i$** | Direction | $Z \to Q$ | $Z \to G$ | $G \to F$ | $G \to Z$ | 1.09 | 240 |
| | Calculation | 0.25 | 1.14 | 0.52 | 0.99 | | |
| | Average | 0.70 | | 0.76 | | | |
| **(x) $C_{O3c}V_{O3c}$** | Direction | $G \to F$ | $G \to Z$ | $G \to F$ | $G \to Z$ | 0.22 | 91 |
| | Calculation | 0.33 | 1.46 | 0.24 | 0.16 | | |
| | Average | 0.90 | | 0.20 | | | |
| **(xii) $C_i V_{O4c}$** | Direction | $G \to F$ | $G \to Z$ | $Q \to F$ | $Q \to Z$ | 6.01 | 521 |
| | Calculation | 1.39 | 0.87 | 11.89 | 1.70 | | |
| | Average | 1.13 | | 6.80 | | | |

### 3.3.1 Dielectric function and dielectric constant

The complex dielectric function $\varepsilon(\omega)$ is the most fundamental function in optics on which all the other optical properties are dependent as expressed in the supporting information formulae. By inspection of the optical properties in **Figure 1S (a,b)** and **2S**, it is unequivocal that m-$ZrO_2$ and the short-listed defected structures reveal almost perfect isotropic behaviour in the dielectric function along the three spatial dimensions. From the real part of the dielectric function, the static dielectric constant $\varepsilon_1(0)$ defines the permittivity of the material that is favourable to be high for solar water splitting purposes as this may lead to lowering $E_{ex}$; leading to improved charge carrier extraction efficiency. All the short-listed defective structures possess slightly higher $\varepsilon_1(0)$ values (4.52-5.45 F/m) than m-$ZrO_2$ (4.13 F/m). For the imaginary part of the dielectric function $\varepsilon_2(\omega)$, $\varepsilon_2(0)$ reflects light-matter interaction between the incident photons and electrons which in turn elaborate the light absorption capability. In our case, all the structures showed no improvement in $\varepsilon_2(0)$ in comparison with the pristine case. Moreover, the main peaks in $\varepsilon_2(\omega)$ shown in **Figure 1S** and **2S** indicate the possible electronic transitions across the main and intermediate bands, so coupling the analysis of $\varepsilon_2(\omega)$ with the absorption function $\alpha(\omega)$ provides sharper insights.

### 3.3.2 Absorption coefficient

All the defective structures enhanced the light absorption with different degrees due to introduction of intermediate bands with different positions within m-ZrO$_2$ bandgap. As displayed in **Figure 1S** and **2S**, the $\varepsilon_2(\omega)$ of the defective structures manifested the presence of peaks within the range of 0-9.5 eV ((x) C$_{O3c}$V$_{O3c}$ offered the highest peak) substantiating the evidence of enhanced absorption and the existence of more electronic transitions relative to pristine case which lacked any peaks in the aforementioned range. To this end, the intermediate bands minimized the bandgap; allowing for more incident photons of less energy within the visible region to get absorbed by the material contributing in the photogeneration of charge carriers. As shown in **Figure 4**, $\alpha(\omega)$ of m-ZrO$_2$ vanishes near 300 nm (ultra-violet region) while the defective structures sustained the absorption within the visible region.

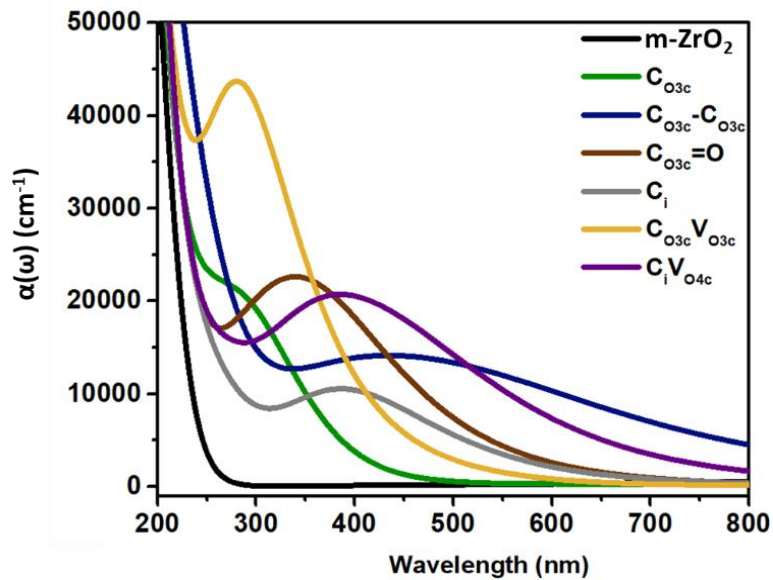

**Figure 4** Absorption coefficient of the short-listed

### 3.3.3 Optical conductivity

The real $R(\sigma)$ and imaginary $Im(\sigma)$ parts of the optical conductivity represent the in-phase current and out-phase inductive currents, respectively. The in-phase current is usually accompanied by liberation of heat energy due to the resistance [81]. In **Figure 1S (c)** and **3S**, the in-phase current dominates over the range 0-34.5 eV (resistance domain) then, the out-phase current takes over after 34.5 eV (induction domain) at which neither energy from the electric field is absorbed nor liberation of heat from the crystal occurs. In the region of 0-5 eV of **Figure 1S (c)** and **3S**, peaks in both real and imaginary parts exist whose positions are consistent with the new emergent bandgaps for each defective structure.

### 3.3.4 Optical reflectivity

The reflectivity spectrum $R(\omega)$ in **Figure 4S (a)** exhibits three peaks at 12, 24, and 36 eV. With the aid of PDOS, the lowest frequency peak signifies the interband transition between the hybridized states within the conduction band while the highest peak indicates the band-to-band transition from the VBM states to the CBM states. Furthermore, the reflectivity at infinite length $R(0)$ is around 0.15 whereas the reflectivity dies out for frequencies more than 50 eV. It is noteworthy that (x) $C_{O3c}V_{O3c}$ offered the lowest average reflectivity.

### 3.3.5 Energy loss function

The energy loss function $L(\omega)$ quantifies the energy lost due to a fast-traversing electron within the material. In the $L(\omega)$ spectrum in **Figure 4S (b)**, the peaks at 14.3, 24.5, and 36.5 eV indicates the characteristic plasma frequencies rather than being interband electronic transitions because $\varepsilon_1(\omega)$ at each peak is equal to zero confirming the occurrence of plasma oscillations [82].

**3.3.6 Refractive index and extinction coefficients**

The refractive index in **Figure 4S (c)** demonstrates nearly the same three peaks at 6.3, 19.2, 30.6 eV while a peak for each defective structure is present within the 0-5 eV range attributed to the various intermediate band positions of the defective structures within the m-ZrO$_2$ bandgap. The static refractive index $n(0)$ is in the range of 2-2.3 for all the defective structures and m-ZrO$_2$. The extinction coefficient $k(\omega)$ describes the absorption loss upon the propagation of an electromagnetic wave through the material. In **Figure 4S (d)**, $k(\omega)$ exhibits three peaks at 10.15, 21.5, and 32.1 eV. For low energy photons, $k(\omega)$ has zero values within the range of the bandgap of each structure. For instance, $k(\omega)$ remains zero from 0- 5 eV which is nearly the bandgap of pristine m-ZrO$_2$. Also, for each defective structure, there is a peak before 5 eV marking the intermediate band position within the bandgap of m-ZrO$_2$.

**4. Conclusions**

In pursuance of the holy-grail overall water splitting, the materials community is striving to design and tune novel efficient electrode materials to uphold this endeavour. In this study, the first-principles calculations embodied in DFT was employed to investigate the thermodynamic, electronic, and optical properties of various carbon doping scenarios in m-ZrO$_2$. The defect formation energies $(E_D^f)$ revealed that elevating the temperature promotes and facilitates the formation of carbon defects. Moreover, it was observed that C$_{O3c}$V$_{O3c}$, and C$_i$ V$_{O4c}$ had the lowest formation energy. Concerning the binding energies ($E_b$), all the complex defects were confirmed to be stable. The analysis of the electronic structure coupled with the absorption coefficient defined the new emerging bandgaps where $C_{O3c}V_{O3c}$ showed the narrowest bandgap (1.9 eV). Furthermore, the band edge positions against the redox potentials of water species elucidated that all the studied defective structures can serve as photoanode. Albeit, $C_{O3c}V_{O3c}$ being the best. It is noteworthy that $C_{O3c}$ was the only defective structure that exhibited the slight straddling of the redox potentials of water species. Upon examining the exciton binding energy ($E_{ex}$), it was noticed that $C_{O3c}V_{O3c}$ possessed the least $E_{ex}$. By the assessment of charge carrier recombination rate via computing the ratio between $m_e^*$ and $m_h^*$ denoted by D, $C_{O3c}$-$C_{O3c}$, $C_{O3c}V_{O3c}$, and $C_iV_{O4c}$ demonstrated the least recombination rate. It is worth mentioning that $C_{O3c}V_{O3c}$ had the lowest $m_e^*$ which implies high mobility and fast charge transfer. The optical properties including dielectric function, absorption coefficient, conductivity, reflectivity, energy loss function, refractive index, and extinction coefficients were studied for the defective structures along with m-$ZrO_2$. Importantly, all the defective structures enhanced the light absorption to different extents. Through the analysis of optical properties, with the exception of the absorption coefficients, the optical properties of the m-$ZrO_2$ and the short-listed defective structures nearly share the same trend with slight variations associated with the different positions of the intermediate bands and new emerging bandgaps for each structure. Nonetheless, $C_{O3c}V_{O3c}$ offered the highest static dielectric constant $\varepsilon_1(0)$, the lowest average reflectivity as well as enhanced light absorption. Thereby, this study highly recommends $C_{O3c}V_{O3c}$ defected m-$ZrO_2$ to be pushed forward for experimental test-bed validations as a photoanode material of tremendous potential for PEC water splitting applications.

**Conflicts of interest**

There are no conflicts to declare.

**Acknowledgements**

We acknowledge the financial support by the Arab-German Young Academy of Sciences and Humanities (AGYA).

# Supporting Information

**Unveiling the role of carbon defects in the exceptional narrowing of m-ZrO$_2$ bandgap for enhanced photoelectrochemical water splitting performance**

*Ahmed H. Biby[a], Sarah A. Tolba[a] and Nageh K. Allam[a*]*

The complex dielectric function (ε) can be described as:

$$\varepsilon = \varepsilon_1 + i\varepsilon_2 = N^2$$

where $\varepsilon_1$ and $\varepsilon_2$ are the real and imaginary parts of the dielectric constant, respectively. The calculation of the imaginary part of the dielectric function is estimated using the following relationship:[81]

$$\varepsilon_2(\omega) = \frac{2e^2\pi}{\Omega\varepsilon_0} \sum_{k,v,c} |\langle \varphi_k^c | H' | \varphi_k^c \rangle|^2 \delta(E_k^c(\vec{k}) - E_k^v(\vec{k}) - \hbar\omega) \quad (1)$$

Where $\Omega$ is the unit cell volume, $\hbar\omega$ is the photon energy, $H'$ is the matrix element for the electromagnetic perturbation added to the normal Hamiltonian taken between the valence and conduction band Bloch states at wave vector $(\vec{k})$, and the $\delta$-function is the energy conservation at $(\vec{k})$. In particular, the imaginary part is calculated first, from which the real part $\varepsilon_1(\omega)$ can be obtained by the Kramers–Kronig transform, using the fact that the dielectric constant describes a causal response. The dielectric constant $\varepsilon(\omega)$ is a function of the frequency $(\omega)$, classifying ε into the electronic contribution part ($\varepsilon_{\omega\to\infty}$), and the lattice vibrational contribution part ($\varepsilon_{\omega=0}$), or the optical and static molecular polarizability. Based on the calculated dielectric constants (refractive index n(ω), extinction coefficient k(ω), absorption coefficient α(ω), reflectivity R(ω), and the energy-loss spectrum L(ω)), the other optical properties can then be obtained using the following relationships:

$$n(\omega) = \frac{\left[\sqrt{\varepsilon_1^2(\omega)+\varepsilon_2^2(\omega)}+\varepsilon_1(\omega)\right]^{0.5}}{\sqrt{2}} \quad (2)$$

$$k(\omega) = \frac{\left[\sqrt{\varepsilon_1^2(\omega)+\varepsilon_2^2(\omega)}-\varepsilon_1(\omega)\right]^{0.5}}{\sqrt{2}} \quad (3)$$

$$\alpha(\omega) = \sqrt{2}\omega\left[\sqrt{\varepsilon_1^2(\omega)+\varepsilon_2^2(\omega)}-\varepsilon_1(\omega)\right]^{0.5} \quad (4)$$

$$R(\omega) = \left|\frac{\sqrt{\varepsilon_1(\omega)+j\varepsilon_2(\omega)}-1}{\sqrt{\varepsilon_1(\omega)+j\varepsilon_2(\omega)}+1}\right|^2 \quad (5)$$

$$L(\omega) = \frac{\varepsilon_2(\omega)}{\left[\varepsilon_1^2(\omega)+\varepsilon_2^2(\omega)\right]} \quad (6)$$

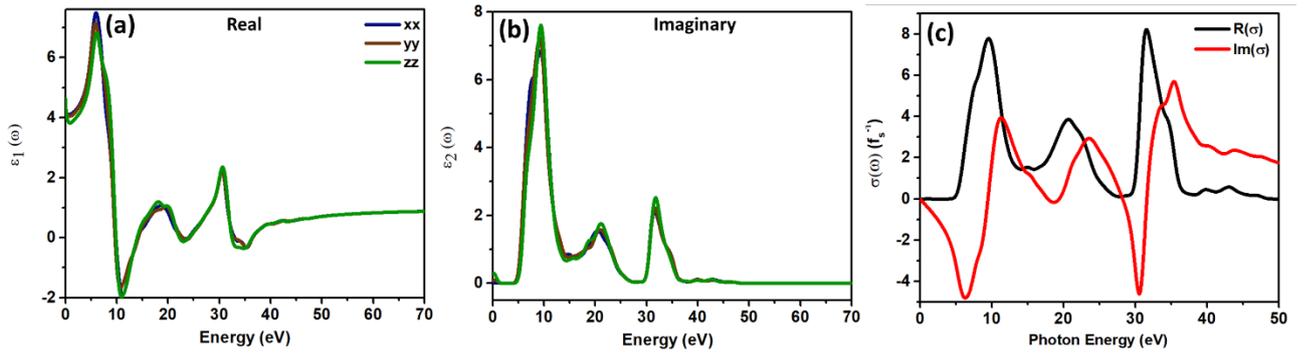

**Figure 1S** Optical properties of m-ZrO$_2$ **(a-b)** The real and imaginary parts of the diagonal components of the dielectric constants of the defected structures respectively, and **(b)** optical conductivity.

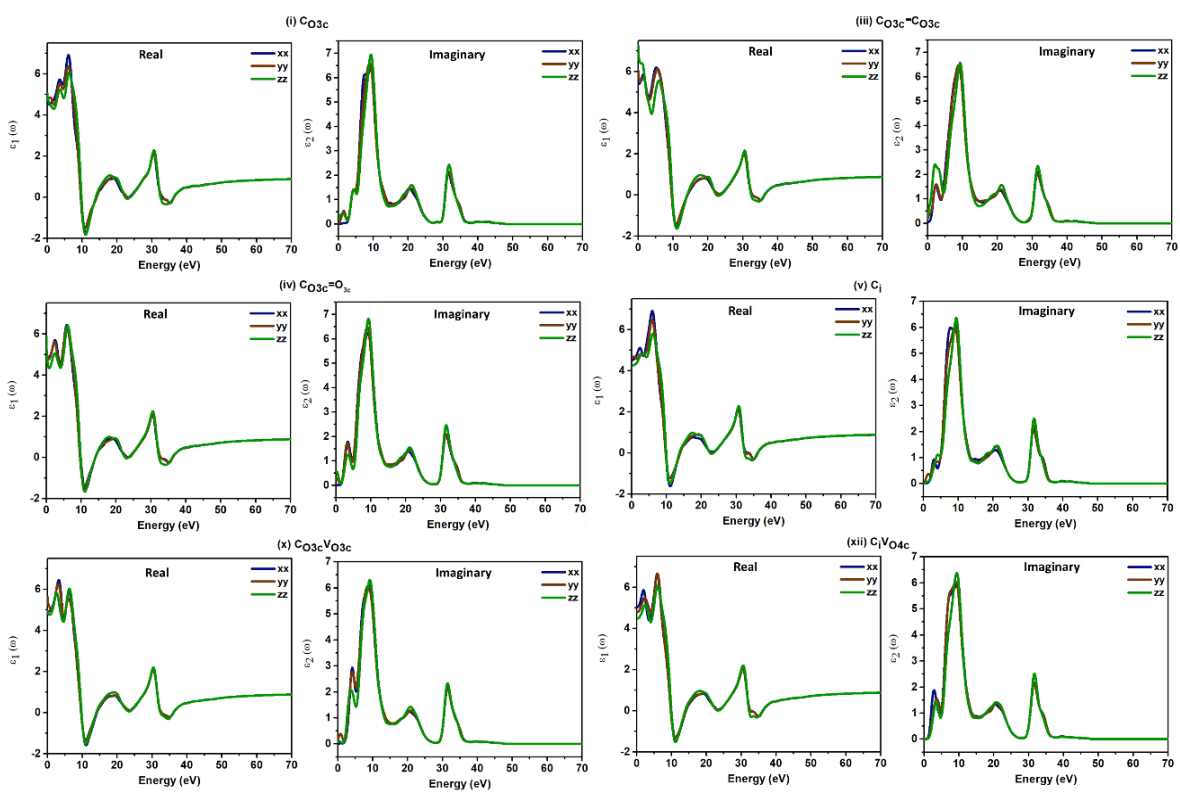

**Figure 2S** The real and imaginary parts of the diagonal components of the dielectric constants of the defected structures.

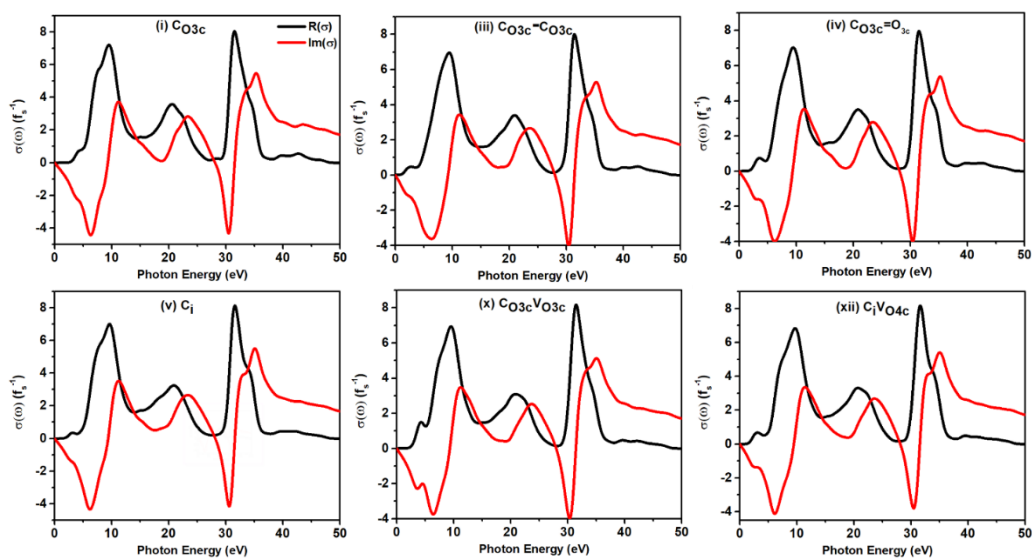

**Figure 3S** Optical conductivity σ(ω) of the defected structures.

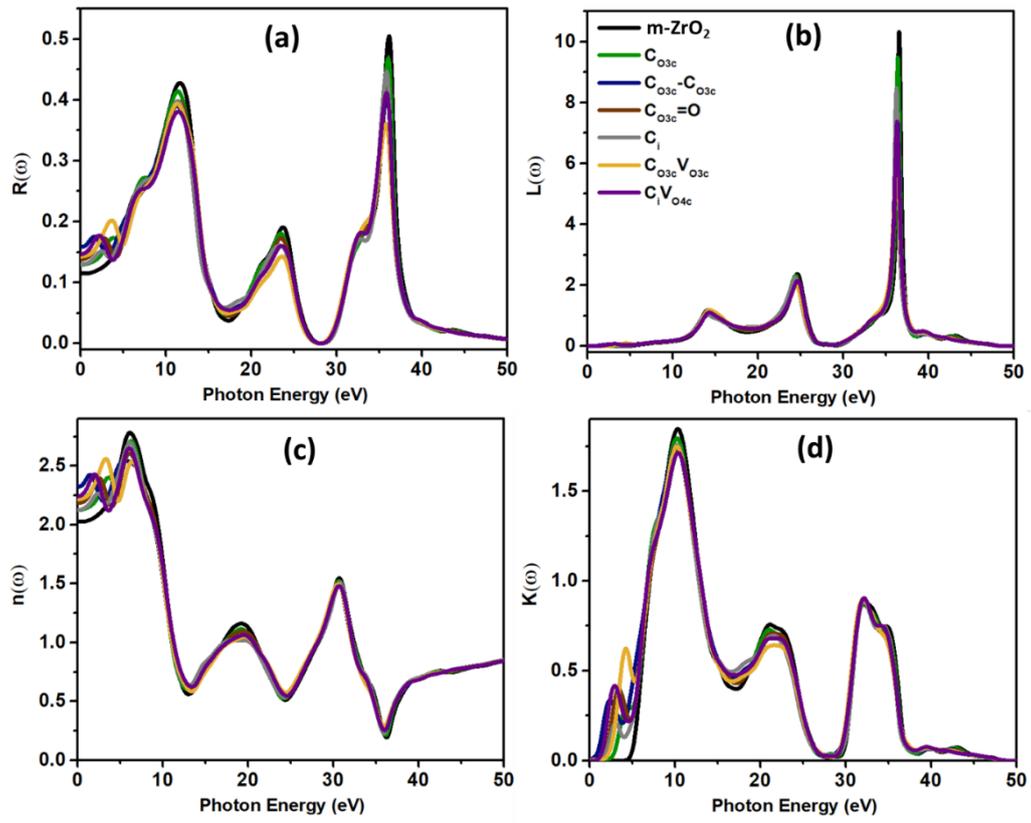

**Figure 4S** Calculated optical properties of the defected structures **(a)** reflectivity R(ω), **(b)** energy loss function L(ω), **(c)** refractive index n(ω), and **(d)** extinction coefficient K(ω).